\documentclass[%
 reprint,
showpacs,
 amsmath,amssymb,
 aps,
prb,
]{revtex4-1}

\usepackage{graphicx}
\usepackage{dcolumn}
\usepackage{bm}

\begin{document}

\title{Far infrared cyclotron resonance and Faraday effect in Bi$_2$Se$_3$}

\author{A. B. Sushkov}
\thanks{sushkov@umd.edu}
\author{G. S. Jenkins}
\author{D. C. Schmadel}
\author{N. P. Butch}
\author{J. Paglione}
\author{H. D. Drew}
\affiliation{Materials Research Science and Engineering Center and Center for Nanophysics and Advanced Materials, University of Maryland, College Park, Maryland 20742}
\date{\today}


\begin{abstract}
The complex Faraday angle and transmission for the topological insulator Bi$_2$Se$_3$ were measured in the far infrared in magnetic fields up to 8~T and analyzed in terms of a simple Drude-Lorentz magneto-plasma dielectric model.
Bulk carriers dominate the optical response.
The bulk carriers are electrons as we determined from the sign of the Faraday angle.
We obtain the bulk band edge cyclotron mass $m_{cb}=0.16 \pm 0.01m_e$ and free electron concentration in the 10$^{17}$ range. 
Electron-phonon interaction effects were found to be weak.

\end{abstract}

\maketitle
\section{Introduction}
It was shown recently theoretically~\cite{Fu2007,Zhang2009} and confirmed experimentally by ARPES~\cite{Hsieh2009,Xia2009} that  Bi$_2$Se$_3$ and Bi$_2$Te$_3$ have an odd number (one) of Dirac cone surface states in the unit cell and thus are nontrivial topological insulators. 
This makes them distinct from graphene which also has Dirac electrons but an even (two) number of cones per unit cell. 
Predictions of several novel properties have generated a great interest in these materials.~\cite{Fu2007,Zhang2009,Hsieh2009} 

Although the optical properties of Bi$_2$Se$_3$ have been studied in the past, only recently it has been probed for unusual far infrared magneto-optical response.~\cite{LaForge}
We have grown Bi$_2$Se$_3$ crystals with low free electron concentration which enabled observation of the  cyclotron resonance and Faraday rotation below the strong low frequency polar phonon. 
In this paper, we report results of these measurements together with the model calculations. 

\section{Experimental}
Single crystals of Bi$_2$Se$_3$ were grown as described in Ref.~\onlinecite{Nick}, and we used a sample from batch iv. 
A 2$\times$2$\times$0.014~mm flake was exfoliated and flattened on a firm surface.
The sample thickness was obtained from etalon oscillations and confirmed by electron microscope images of the sliced flake.  
Infrared data were obtained on a Bomem DA3 Fourier-transform (FTIR) spectrometer and an optically pumped far infrared laser at a frequency of 42.3~cm$^{-1}$. 
The samples were placed either in an optical cryostat for $B=0$ measurements or in an optical split-coil magnet for measurements in magnetic fields up to 8~T in the Faraday geometry: $k_{light} || B$ and $ B || c$-axis. 
Our technique for the complex Faraday angle measurements in the far infrared is described in Refs.~\onlinecite{Jenkins2010,Jenkins2007}. 
  
\section{Model}
The experiments reported here differ from conventional cyclotron resonance in semiconductors because the relatively high carrier density and low lying phonon leads to strong magneto-plasma effects. 
This results in a very low transmission near the plasma frequency and multiple reflections inside the flake in transparent regions (etalon effect) which produces a complicated optical response. 
Therefore, it is necessary to model the magneto-optical response in order to obtain the material parameters. 
We model our experiment as transmission and reflection of a slab with bulk optical constants and (possibly) two identical 2D layers with surface optical constants on both surfaces of the slab. 
For our free carrier concentration level, the Fermi energy is less than 20~meV above the bottom of the lowest conduction band. 
This is the only filled band because, according to Ref.~\onlinecite{Kulbachinskii}, the second conduction band minimum is 40~meV higher. Since $\hbar \omega$ and $E_F$ are small compared with the band gap $\approx$300~meV, a parabolic dispersion of the bulk electrons is assumed. 
Therefore, experiment gives essentially the band edge mass.  

In this case, we can use a Drude magneto-optical response of the free electron gas (1st term in Eq.~\ref{eps}). 
The corresponding complex dielectric function for the bulk is: 
\begin{equation}
\epsilon^{\pm}(\omega) = - \frac{\omega_{pb}^2}{\omega (\omega \pm \omega_{cb} -\imath \gamma_b)}
+\sum_j\frac{S_{j}}{\omega_{j}^2-\omega (\omega - \imath \gamma_j)}
+ \epsilon_\infty
\label{eps}
\end{equation}
where  $j$ enumerates the phonon oscillators, 
$\omega_{j}$ is the TO phonon frequency, 
$S_j$ is the phonon spectral weight, 
a subscript $b$ refers to bulk electrons, 
$\epsilon_\infty$ is the dielectric constant at higher infrared frequencies,
$\omega_{pb}$ is the unscreened plasma frequency for free electrons: $\omega_{pb}^2=4\pi N_be^2/m_b^*$, where $N_b$, $e$ and $m_b^*$ are concentration, charge, and an effective mass of electrons;  
$\gamma_b$ and $\gamma_j$ are the damping rates; 
$\omega_{cb} = e B / (m_{cb} c)$ is the cyclotron frequency, where $B$ is magnetic field, $m_{cb}$ is bulk cyclotron mass, and $c$ is speed of light. 
The $\pm$ sign refers to the right and left circular polarized electromagnetic modes. 
For a fixed (laser) frequency, we use a quarter-wave plate to obtain circular polarized light. 
In case of the far infrared spectra, we used linear polarized light thus measuring an average of two circular polarizations.
For a general polarization state, measured transmission is $T = (0.5 + \beta) T^+ + (0.5 - \beta) T^-$, where $-0.5 \leq \beta \leq 0.5$.

The surface states of a topological insulator are 2D Dirac quasiparticles that split into Landau levels in an applied $B||c$ magnetic field : 
\begin{eqnarray}
E_n = sgn(n)\sqrt{n}\hbar \omega_{cs} , \nonumber \\ 
\omega_{cs} = v_F \sqrt{2eB/\hbar}
\label{LL}
\end{eqnarray}
where integer $n$ enumerates energy levels, $e$ and $v_F$ are charge and Fermi velocity of the Dirac particles.  The full band dispersion for Bi$_2$Se$_3$ has been recently reported in Ref.~\onlinecite{Liu2010}.  
The optical dipole selection rules allow transitions $|n| \rightarrow |n|+1$ for the cyclotron resonance active mode of circularly polarized light. 
The linear dispersion leads to the square root dependence on $n$ and $B$. 
However, in our sample the chemical potential $\mu$ is $\approx 10$~meV above the bottom of the conduction band which, in turn, is $\approx 200$~meV above the Dirac point. 
Therefore, we find $n\approx 25$ for 8~T magnetic field. 
At these values of $n$, we can use an approximate formula 
\begin{equation}
E_{n+1} - E_{n} \approx \hbar\frac{eBv_F^2}{cE_n} = \hbar \omega_{cs}
\end{equation}
and the cyclotron resonance becomes semiclassical.  
The combination $E_n/v_F^2$ plays the role of the cyclotron mass $m_{cs}$ for surface states. 
Therefore, we model the optical response of the surface states as a 2D free electron gas. 

We use conventional expressions for transmission and reflection of a slab (medium 2, bulk) between media 1 and 3 (both vacuum here) for light propagating from 1 to 3 with the following modifications. 
The presence of surface states at the interfaces 1-2 and 2-3 changes the complex Fresnel transmission and reflection coefficients:
\begin{eqnarray}
t_{12}^{\pm} = \frac{2n_1}{n_1 + n_2^{\pm} + y^{\pm}}, 
r_{12}^{\pm} = \frac{n_1 - n_2^{\pm} - y^{\pm}}{n_1 + n_2^{\pm} + y^{\pm}}, \nonumber \\ 
r_{21}^{\pm} = \frac{- n_1 + n_2^{\pm} - y^{\pm}}{n_1 + n_2^{\pm} + y^{\pm}}, \\
t_{23}^{\pm} = \frac{2n_2}{n_3 + n_2^{\pm} + y^{\pm}}, 
r_{23}^{\pm} = \frac{n_2^{\pm} - n_3 - y^{\pm}}{n_2^{\pm} + n_3 + y^{\pm}}, \nonumber 
\label{tr}
\end{eqnarray} 
where $n_1=n_3=1$, $n_2^{\pm}=\sqrt{\epsilon^{\pm}(\omega)}$ are refractive indices, and surface admittance $y^{\pm}$ is
\begin{equation}
y^{\pm} = \frac{\omega_{ps}}{\gamma_s - \imath (\omega \pm \omega_{cs})} , 
\label{y}
\end{equation} 
where $\omega_{ps} = 4\pi N_s e^2/(m^*_{s} c)$ is a 2D plasma frequency, a subscript $s$ refers to the surface states.
Using standard expressions for the total complex transmission $t$ and reflection $r$ amplitude coefficients of a slab 
\begin{equation}
t^{\pm} = \frac{t_{12}^{\pm} t_{23}^{\pm} \phi}{1- r_{21}^{\pm} r_{23}^{\pm} \phi^2},
r^{\pm} = r_{12}^{\pm} + \frac{t_{12}^{\pm} t_{21}^{\pm} r_{23}^{\pm} \phi^2}{1- r_{21}^{\pm} r_{23}^{\pm} \phi^2}, 
\label{tpm}
\end{equation} 
where $\phi = \exp{(\imath 2\pi n_2^{\pm} \omega d)}$ and $d$ is a sample thickness, 
we calculate transmission $T^{\pm}=|t^{\pm}|^2 n_3/n_1$ and reflection $R^{\pm}=|r^{\pm}|^2$ intensity coefficients. 
We define the complex Faraday angle $\theta_F$ via $\tan \theta_F = t_{yx}/t_{xx}$, where standard transformation between circular and linear polarization bases $t^{\pm}= t_{xx} \pm \imath t_{xy}$ is assumed. 

We estimate $\omega_{ps}$ using the ARPES data~\cite{Hsieh2009,Xia2009} and expressions for 2D free Fermi gas (one spin per state): $k_F^2 = 4\pi N_s$. 
For $E_F \approx 10$~meV, the surface states are below $k_{F}=0.08$~$\AA^{-1}$ which gives $N_s \approx  5\times 10^{12}$~cm$^{-2}$ and  $\omega_{ps} \approx 20$~cm$^{-1}$ assuming $m^*_{s}=0.15m_e$. 

\section{Experimental results and discussion}
Zero field far infrared results for this sample were reported earlier.~\cite{Nick}
Set \#1 shown in Table~1 is the result of a simultaneous fit of zero field reflectivity and transmission. 
Other sets of fit parameters shown in Table~1 model results from different magnetic field experiments discussed below. 
Data of experiment \#2 were fitted by the model of two crystals with parameters \#2-1 and \#2-2 and effective areas 82~\% and 18~\%, correspondingly. 
We understand variation of the bulk concentrations from one experiment to another and within experiment \#2 as a consequence of nonuniform carrier concentration of the sample. 
Characteristic size of the regions with different concentrations should be on the order of the infrared laser focal spot ($\approx 1$~mm). 
In FTIR measurements, the light spot was larger than the sample aperture. 
We also cannot rule out a possible increase of the effective bulk concentration with time due, for example, to exposure of the flake to air during the few months separating measurements \#1 and \#4.
\begin{table}
\caption{Fit parameters for Bi$_2$Se$_3$ flake and calculated values (last 3 columns).}
\begin{ruledtabular}
\begin{tabular}{lcccccc}
 Set\footnotemark[1] & $\omega_{pb}$ & $\gamma_b$ & $m_{cb}$ & $\omega_{cb}$ (8 T) & $N_b$\footnotemark[2] & $E_F$\footnotemark[2]\\
 (\#) & (cm$^{-1}$)      & (cm$^{-1}$)   & ($m_e$)    & (cm$^{-1}$)            & ($10^{17}$ cm$^{-3}$) & (meV) \\
\hline
 1  (FTIR) & 382            & 8 &  0.15    & 50                   & 2.3                 & 8  \\
 2-1 (laser)   & 385            & 7.5 &  0.16    & 48                  & 2.6                 & 8 \\
 2-2 (laser) & 750            & 7.5 &  0.16    & 48                  & 9.9               & 20 \\
 3 (FTIR)  & 465            & 8 &  0.16    & 48                   & 3.8                 & 11 \\
 4 (laser)  & 520            & 8 &  0.17    & 44                   & 5.1                 & 12 \\
\end{tabular}
\end{ruledtabular}
\footnotetext[1]{In chronological order.
Fixed phonon parameters are 
$\omega_1=67$~cm$^{-1}$, $S_{1}=3.94\times 10^5$~cm$^{-2}$, $\gamma_1= 5~$cm$^{-1}$,  
$\omega_2=134$~cm$^{-1}$, $S_2=7056$~cm$^{-2}$, $\gamma_2= 2$~cm$^{-1}$, and
$\epsilon_\infty = 25.6$.} 
\footnotetext[2]{$m_b^* = m_{cb}$}
\end{table}

\begin{figure}
\includegraphics[width=\columnwidth]{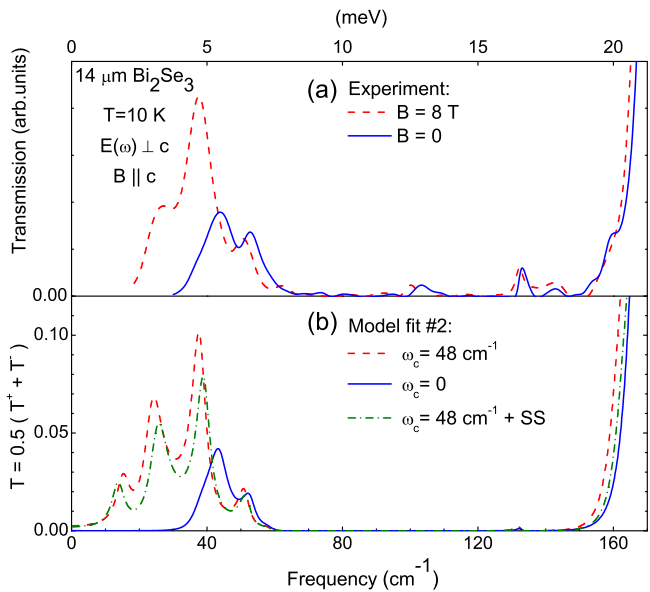}
\caption{\label{tra}(Color online) Transmission of Bi$_2$Se$_3$ flake for unpolarized light in magnetic field $B||c$: (a) measured and (b) calculated;  green dash-dot curve is calculated with surface states (SS) included. } 
\end{figure}

Figure \ref{tra} shows the measured and calculated transmission spectra of the Bi$_2$Se$_3$ flake. 
Oscillations of the transmission in zero field appearing below 60~cm$^{-1}$ are due to the etalon effect within the transparency band between the plasma frequency and the TO phonon frequency.
In non zero fields, the magneto-plasma dielectric function (1) changes due to the cyclotron resonance, and the etalon features for left and right circular polarization split in frequency and change their amplitudes.  
The material becomes more transparent in magnetic field because the $T^+$ (cyclotron resonance inactive) polarization becomes less absorbing. 
In the fit \#3 shown in fig.~\ref{tra}(b), we match frequency positions of the main maxima in the experimental and model transmissions. 
The green dash-dot curve in fig.~\ref{tra}(b) shows how much the transmission changes if surface states are added with an exaggerated strength $\omega_{ps} = 100$~cm$^{-1}$ ($\gamma_s=8$~cm$^{-1}$, $m_{cs} = m_{cb} = 0.16m_e$). 
The curve calculated for the estimated ARPES value of $\omega_{ps} = 20$~cm$^{-1}$  is almost indistinguishable from the $\omega_{ps} = 0$ red dashed curve. 

\begin{figure}
\includegraphics[width=\columnwidth]{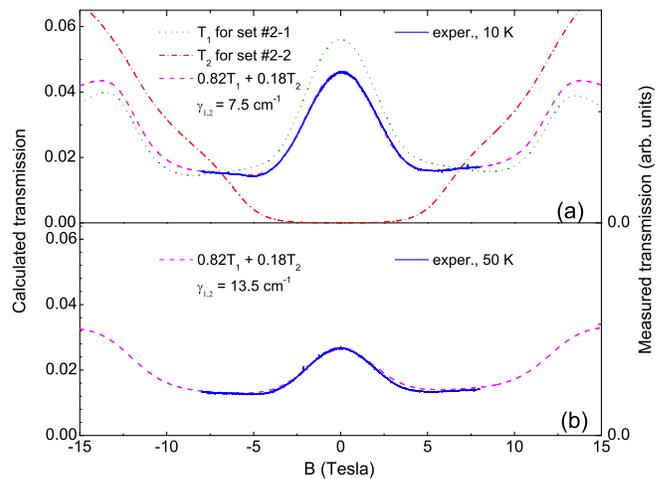}
\caption{\label{fit1}(Color online) Measured (blue solid lines) and calculated (bulk only) transmission of nearly unpolarized light ($\beta= -0.02$) for Bi$_2$Se$_3$ flake at the fixed laser frequency 42.3~cm$^{-1}$. 
(a) The calculated transmission for the regions of the flake with low carrier concentration (green dotted curve), high concentration (red dash-dotted curve), and a weighted admixture of the two which fits the 10~K data (pink dashed curve).
(b) The same model and parameters used in (a) fit the 50~K data provided the scattering rate is raised to $\gamma_1=\gamma_2=13.5$~{cm}$^{-1}$ (pink dashed curve). } 
\end{figure}

We have also measured unpolarized magneto-transmission of the sample at the fixed laser frequency 42.3~cm$^{-1}$. 
Figure \ref{fit1} shows measured and calculated transmission spectra of the same Bi$_2$Se$_3$ sample as a function of applied magnetic field. 
The upturn of the experimental curve above 5~T cannot be reproduced by our simple model with one concentration either with or without surface states. 
However, we can fit the entire experimental curve by the weighted average of two independent sets of bulk parameters, simulating inhomogeneity of the free carrier concentration, described by sets \#2-1 and 2-2. 
Including different thicknesses associated with each set of bulk parameters to concurrently simulate thickness variations can not significantly reduce the concentration difference between  sets \#2-1 and 2-2 if the thicknesses are limited to  $14 \pm 4$~microns as measured by scanning electron microscope. 
Based upon these simple models, it appears that concentration inhomogeneity may cause  deviations from the simple line shape expected from a single uniform bulk response. 
A distribution of concentrations will cause a distribution of plasmons (zero crossings of the real part of epsilon) which will affect the optical response in a nontrivial way. 

It should be noted that all measurements are in the free carrier concentration range set by parameters \#2-1 and 2-2. 
The 50~K data is very well fit with the same parameters as fit the 10~K data except $\gamma$ requires an increase from 7.5 to 13.5~cm$^{-1}$. 
This suggests that only the free carrier damping rate changes over this temperature range. At higher temperatures, phonon parameters and effective free carrier concentration may also change.

\begin{figure}
\includegraphics[width=\columnwidth]{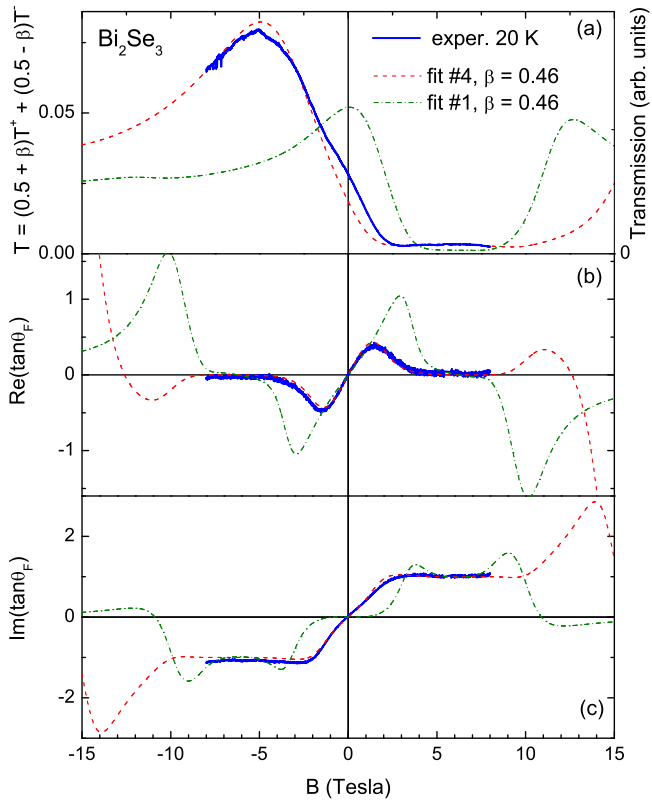}
\caption{\label{fit3}(Color online) (a) Transmission of circular polarized laser light at 42.3~cm$^{-1}$, (b) real and (c) imaginary parts of $\tan \theta_F$ ($\theta_F$ --- complex Faraday angle). } 
\end{figure}

Figure \ref{fit3} shows the transmission of circular polarized light and the complex Faraday angle (Faraday rotation and circular dichroism) together with  a model calculation with one set of parameters (\#4). 
The green dash-dot curves were calculated using fit~\#1 parameters and clearly deviate strongly from the experimental results. 
The qualitative difference in shapes of the transmission curves for figures \ref{fit1} and \ref{fit3}(a) is also understood in terms of carrier concentration inhomogeneities.

The sign of $\omega_c$ in our magneto-optical experiments is determined by the sign of the carrier charge, the direction of magnetic field, the chirality of incident light, and other experimental conditions. 
To avoid any possible ambiguity, a boron doped Si wafer with an anti-reflection coating in which the free carriers are holes was measured as a reference sample. 
The measured Faraday rotation for the Bi$_2$Se$_3$ flake has the same sign as for the Si:B wafer.
However, the sign of Bi$_2$Se$_3$ carriers is opposite to holes because of a sign change of the Faraday response due to an etalon effect in the flake as we see from the model calculations. 
Therefore, the free carriers in our Bi$_2$Se$_3$ sample are electrons.

These results also bear on electron interaction effects.  
While the Coulomb interaction is weak because of the large static screening ($\epsilon_0 \approx 100$), a significant electron-phonon interaction may be expected in this strongly polar material. 
In either case, the carriers scattering rates and effective masses reflect the significance of the many body interaction effects. 
Electron-electron scattering is expected to have $T^2 + \omega^2$ dependence from Coulomb scattering and an onset of scattering for $\omega > \omega_{LO}$ due to LO phonon emission. 
In addition, the optically measured cyclotron mass differs, in principle, from the value measured from quantum oscillations measurements which includes the self energy corrections to the band mass due to interactions. 
For the optical measurement of cyclotron resonance, the Kohn theorem insures that the resonance is controlled by the band mass provided that the electrodynamics is sufficiently local.~\cite{Kohn,Ando} 
In the Faraday geometry of our experiments, the non locality leads to a Doppler shift of the cyclotron resonance at $\omega = \omega_c - q v_F$, where $q$ is a characteristic electrodynamic wave vector.~\cite{Platzman}  
For this system $q \approx \omega_p/c \approx 10^{10} << \omega_c/v_F$, so that the expected shifts are very small.  
Therefore, the observed cyclotron mass should correspond to the band mass. 
The optically measured $m_{cb} = 0.15 \pm 0.01$. 
The cyclotron mass deduced from SdH measurements on samples from the same batch was $m_{SdH} = 0.15 \pm .01$.  The near equality of these two values implies that mass enhancement from many body interactions is very small in this material.  

\begin{figure}
\includegraphics[width=\columnwidth]{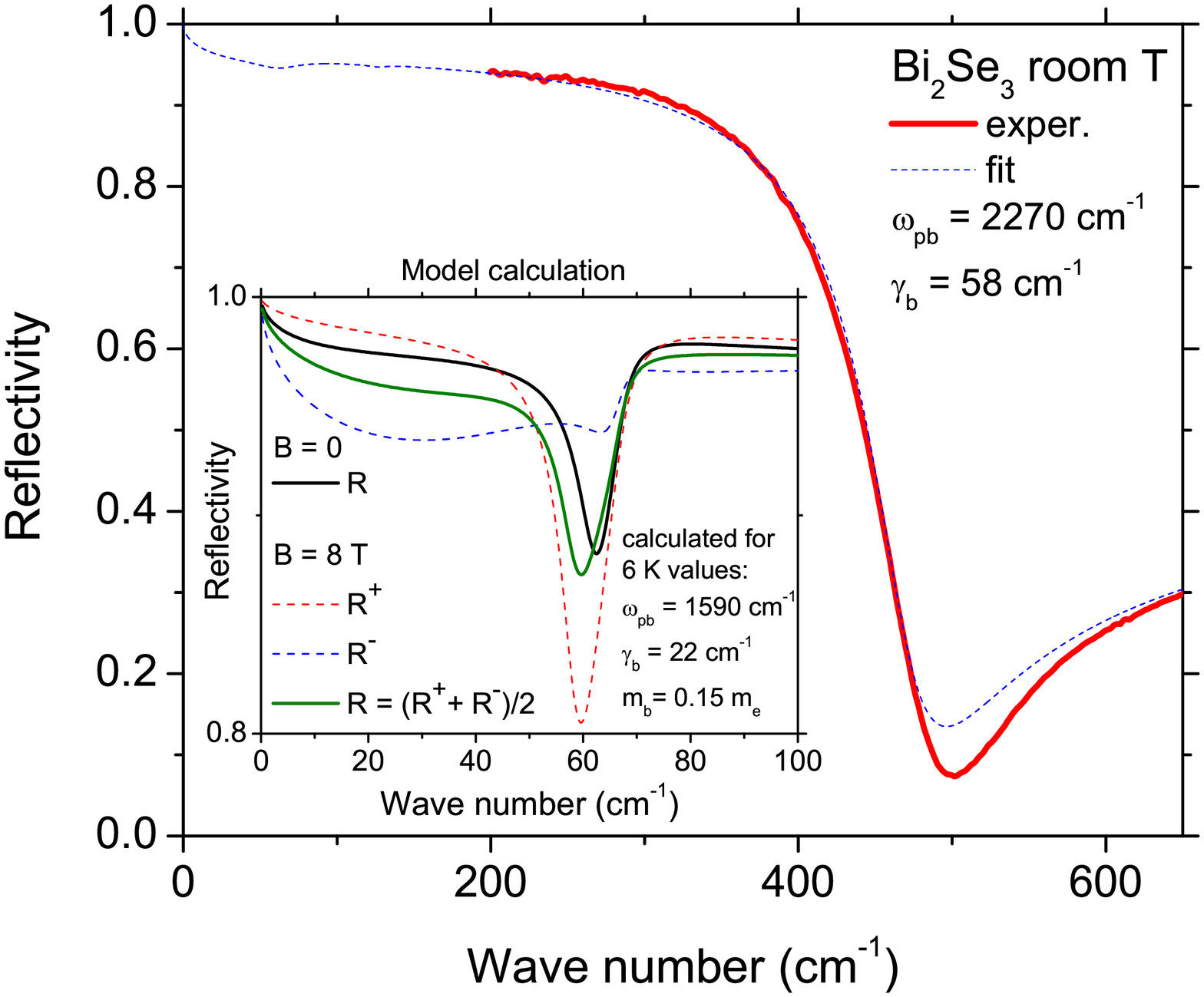}
\caption{\label{el-ph}
(Color online) Reflectivity of a Bi$_2$Se$_3$ crystal with $N_b = 8.5\times 10^{18}$.  Inset: model calculations for the phonon line shape in magnetic field using data of Ref.~\onlinecite{LaForge}. 
Deviations in line shape and an apparent phonon frequency shift from the B=0~T (solid black) curve is due to the differences in the optical responses of left (dashed red) and right (dashed blue) circularly polarized light at fields of 8~T. Linearly or unpolarized light is an equal admixture of the two responses (solid green).  } 
\end{figure}

In figure~\ref{el-ph}, we show the room temperature reflectivity spectrum of a crystal with a higher free carrier concentration ($N_b = 8.5\times 10^{18}$) than the flake. 
The spectrum shows a plasma edge at 500~cm$^{-1}$.  
The plasma edge feature is very sensitive to the electron relaxation rate at this frequency.  
The electronic relaxation rate implied by this spectrum is $\gamma_b = 58$~cm$^{-1}$.  
It is noteworthy that this relaxation rate is relatively small considering that the plasma edge is significantly above the LO phonon frequency which is $\omega_{LO} \approx 150$~cm$^{-1}$.  
This relaxation rate is comparable to the relaxation rate measured at lower frequencies at room temperature so that LO phonon scattering does not add significantly to the relaxation rate.  

Therefore, from these observations, we see that the electron-phonon interaction effects are too small to be observed on the scale of the uncertainty of the effective mass and the impurity scattering rate. 
This surprising conclusion may result from the free carrier screening of the electron-phonon interaction.

It was argued in Ref.~\onlinecite{LaForge} that the electron-phonon interaction is strong based in part on the anomalous line shape of the lowest infrared phonon. 
In fig.~\ref{el-ph} (inset), we show results of our model calculation for this phonon in magnetic field at low temperatures. 
The line shape of the phonon in magnetic field is an average of two very different reflectivities $R^+$ and $R^-$ which results in apparent softening and broadening of the phonon line. 
Our transmission data presented in fig.~\ref{tra} also show insensitivity of the lowest phonon to applied magnetic field. 
Thus, the observations of Ref.~\onlinecite{LaForge} on this phonon at least partially originate in cyclotron resonance physics rather than in electron-phonon and magneto-electric couplings. 

In conclusion, we have measured the bulk cyclotron mass for a low doped Bi$_2$Se$_3$ single crystal which is close to the mass observed in SdH measurements.  
Free carriers are electrons as we determined from the sign of the Faraday angle. 
We have shown that an effective Drude-Lorentz model well describes the magneto-optical data which is dominated by etalon effects in the slab sample. 
No signature of strong electron-phonon interaction is observed.

\begin{acknowledgments}
We thank Shou-Cheng Zhang and Xiao-Liang Qi for fruitful discussions. 
This work was supported in part by the National Science Foundation MRSEC under Grant No. DMR-0520471. 
 N. P.~Butch was supported by CNAM.
\end{acknowledgments}

\end{document}